\documentclass[11pt,a4paper]{article}
\usepackage[noadjust]{cite}

\usepackage[margin=2.8cm,bottom=3.5cm]{geometry}

\usepackage{amsmath, amssymb}
\usepackage{color}
\pdfoutput=1 

\usepackage{graphicx} 
\usepackage{subcaption}
\usepackage[T1]{fontenc}

\title{Wobbling double sine-Gordon kinks}

\author{
        Jo\~ao G. F. Campos \\
        Departamento de F\'isica, Universidade Federal de Pernambuco,\\
        Av. Prof. Moraes Rego, 1235, Recife - PE - 50670-901, Brazil\\
        joao.gfcampos@ufpe.br
            \and
        Azadeh Mohammadi\\
        Departamento de F\'isica, Universidade Federal de Pernambuco,\\
        Av. Prof. Moraes Rego, 1235, Recife - PE - 50670-901, Brazil\\
        azadeh.mohammadi@ufpe.br
}

\begin{document} 
\maketitle

\begin{abstract}

We study the collision of a kink and an antikink in the double sine-Gordon model with and without the excited vibrational mode. In the latter case, we find that there is a limited range of the parameters where the resonance windows exist, despite the existence of a vibrational mode. Still, when the vibrational mode is initially excited, its energy can turn into translational energy after the collision. This creates one-bounce as well as a rich structure of higher-bounce resonance windows that depend on the wobbling phase being in or out of phase at the collision and the wobbling amplitude being sufficiently large. When the vibrational mode is excited, the modified structure of one-bounce windows is observed in the whole range of the model's parameters, and the resonant interval with higher-bounce windows gradually increases with the wobbling amplitude. We estimated the center of the one-bounce windows using a simple analytical approximation for the wobbling evolution. The kinks' final wobbling frequency is Lorentz contracted, which is simply derived from our equations. We also report that the maximum energy density value always has a smooth behavior in the resonance windows. 
\end{abstract}

\section{Introduction}

Solitons, instantons and monopoles for instance, are important solutions of field theories that appear when the configuration of the system at the boundary is topologically nontrivial \cite{rajaraman1982solitons, manton2004topological}. In particular, soliton solutions of relativistic scalar field theories in ($1+1$) dimensions are called kink and antikink. The degeneracy of the potential is essential for their existence. These solutions appear in the description of many physical systems such as polyacetylene \cite{su1979solitons}, Josephson junctions \cite{vachaspati2006kinks}, graphene deformations \cite{yamaletdinov2017kinks}, domain walls in ferromagnets \cite{kardar2007statistical} and Helium-3 \cite{volovik2003universe}.

Two of the most studied models in the field are the sine-Gordon and the $\phi^4$. The sine-Gordon is integrable, giving rise to what is called a true soliton. The $\phi^4$, on the other hand, is non-integrable and exhibits a much wider variety of outcomes. In the former, the collision between the kinks is always elastic, and the only effect of the interaction is a phase shift in the kinks propagation. In the latter, the constituents in a kink-antikink collision may instead reflect inelastically or annihilate. Moreover, it may also happen that the solitons separate after more than one-bounce in what is called resonance windows. This interesting phenomenon has been extensively studied for a few decades and had no compelling quantitative explanation until recently \cite{manton2021kink, manton2021collective}.

Historically, the pioneering works about kink-antikink collisions include  Sugiyama \cite{sugiyama1979kink}, Campbell et al. \cite{campbell1983resonance, peyrard1983kink, campbell1986kink}. and Anninos et al. \cite{anninos1991fractal}. In \cite{sugiyama1979kink}, the author observed that, after the collision, a kink and an antikink annihilate or reflect after a critical velocity. Moreover, the author proposed a reduced model describing the system using collective coordinates. Unfortunately, one of the equations had a typo that propagated in the literature for many years. In the triplet of papers \cite{campbell1983resonance, peyrard1983kink, campbell1986kink} the authors computed the resonance windows for the $\phi^4$ model, a modified sine-Gordon model, and the double sine-Gordon with high precision.
Moreover, they argued that the resonance windows occur due to a resonant energy exchange mechanism between the kink's translational and vibrational modes. Thus, they found analytical expressions for the windows' location and shape and conjectured that higher-order resonance windows exist at the border of the lower-order ones. In \cite{anninos1991fractal} the authors showed that resonance windows form a fractal structure, both numerically and using the reduced system proposed by Sugiyama. Unfortunately, when the typo in the reduced equations was corrected, the qualitative similarity between the reduced and full systems disappeared \cite{takyi2016collective}. To remedy that, the authors of \cite{takyi2016collective} tried to integrate the reduced equations without any approximation but faced a singularity in the equations. Finally, in two recent works, this singularity problem was corrected with a clever choice of collective coordinates \cite{manton2021kink}, and then the reduced equations could reproduce the resonance structure \cite{manton2021collective}. 

There has been a great number of works discussing kink-antikink interactions in various scenarios in the literature. Some of these include the investigation of quasinormal modes in kink-antikink collisions \cite{dorey2018resonant,campos2020quasinormal}, models with BPS preserving defects \cite{adam2019phi, adam2019spectral, adam2019bps, adam2019solvable, manton2019iterated, adam2020kink}, interactions of kinks with fermions \cite{gibbons2007fermions, saffin2007particle, chu2008fermions, brihaye2008remarks, campos2020fermion}, collisions between kinks with long-range tails \cite{khare2019family, christov2019long, manton2019forces, christov2019kink, christov2020kink, campos2020interaction}, collision of kinks with boundaries \cite{arthur2016breaking, dorey2017boundary, lima2019boundary}, multikink scattering \cite{marjaneh2017multi, marjaneh2017high, marjaneh2018extreme, gani2019multi} and collision between kinks in two-component scalar field theories \cite{halavanau2012resonance, alonso2018reflection, alonso2020non}.

The double sine-Gordon model is a compelling non-integrable model which becomes integrable in some limits. Therefore, it is possible to study a gradual transition from integrability to non-integrability. Early works of the double sine-Gordon model studied it as a perturbation of the sine-Gordon model via the inverse scattering method \cite{kivshar1987radiative, malomed1989dynamics,	kivshar1989dynamics}. However, to leading order, a conservative perturbation to the sine-Gordon model still has a trivial kink-antikink collision \cite{kivshar1989dynamics}. Other important works about kink collisions in this model include \cite{campbell1986kink, gani1999kink, gani2018scattering, gani2019multi, simas2020solitary}. An important feature that appears in these works is that the kinks have an inner structure, i.e., it consists of two subkinks, which may be exchanged at collision and form subkink bound states. This phenomenon was also observed in other systems with the kinks with inner structures \cite{zhong2020collision}. Curiously, the double sine-Gordon model may effectively describe some physical systems such as gold dislocations \cite{el1987double}, optical pulses and spin waves \cite{bullough1980double}, pseudo 1-D ferromagnets \cite{rettori1986double} and Josephson structures \cite{alfimov2014discrete}. Despite all that, it has not been sufficiently explored in the literature. It was only recently that the dependence of the critical velocity on the model parameter $R$ was calculated \cite{gani2018scattering}.

Recently Alonso-Izquierdo et al. studied a fascinating problem for kink-antikink interactions \cite{izquierdo2021scattering}. They considered collisions between wobbling kinks, meaning that the kinks have their vibrational mode excited at the collision. A single wobbling kink's behavior has already been studied for the $\phi^4$ model \cite{manton1997kinks, barashenkov2009wobbling, oxtoby2009resonantly}. In particular, it is well known that the wobbling amplitude decreases due to coupling to radiation via the first harmonic \cite{manton1997kinks} or, for some other models, via higher harmonics \cite{romanczukiewicz2018oscillons}. 
The collision between wobbling kinks is important because the successive bounces in resonance windows may be seen as an iteration of such events. Interestingly, the authors showed in \cite{izquierdo2021scattering} that there appear many separate one-bounce windows due to the wobbling.  
Here, we study collisions between wobbling double sine-Gordon kink and antikink. Interestingly, when the wobbling is turned off, the model does not exhibit resonance windows near its integrable limits, despite having a vibrational mode. This seems to be an exception to the resonant energy exchange mechanism. However, when the wobbling is turned on, the resonant structure is gradually recovered. We show that the double sine-Gordon shares many similar features with the $\phi^4$ model, such as one-bounce resonance windows. We are able to approximate the locations of these windows using the simplified analysis of the resonance windows structure of Campbell et al. Furthermore, this can also be visualized by plotting the maximum value of the energy densities similar to the analysis in \cite{gani2019multi, yan2020kink}. The structure of the paper is as follows. In section \ref{model} we give a brief review of the double sine-Gordon models. In section \ref{simul} we show and compare the numerical results of our simulations of kink-antikink collisions with and without wobbling. Finally, in the last section, we give our concluding remarks.
\section{Model}
\label{model}

\begin{figure}[tbp]
\centering
   \includegraphics[width=\linewidth]{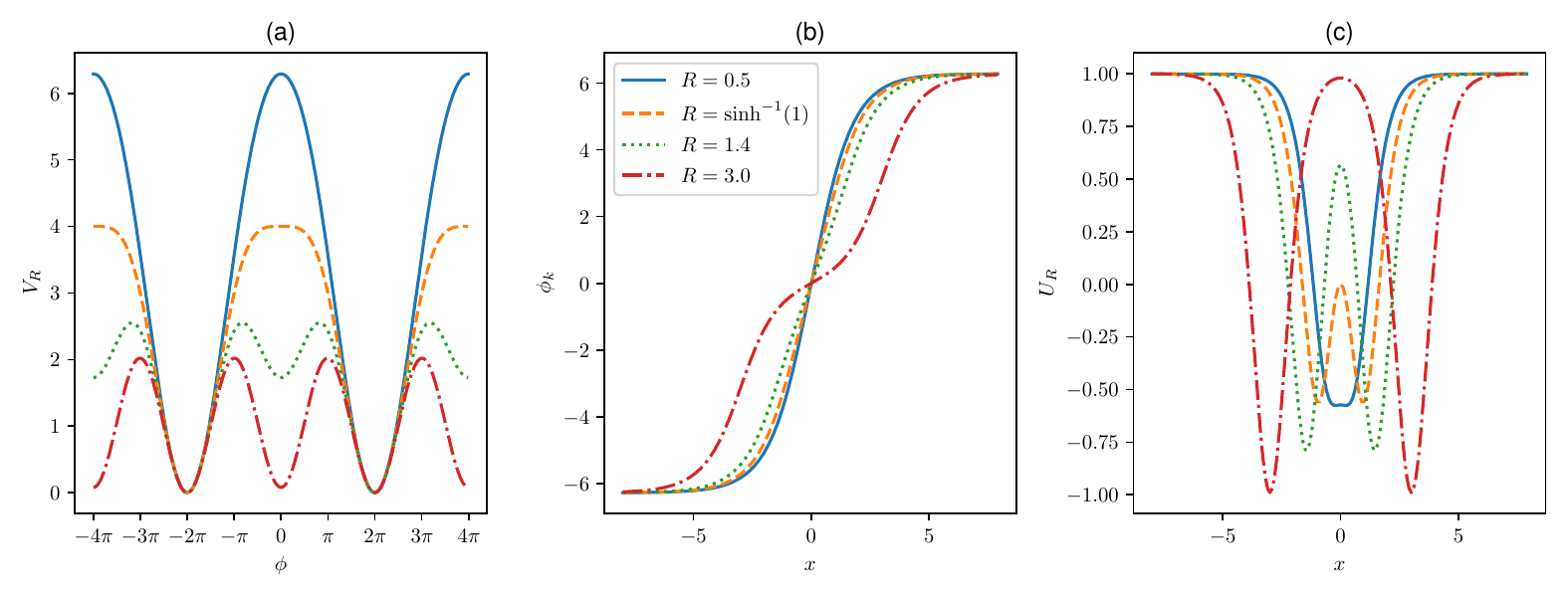}
   \caption{(a) Potential, (b) kink profile, and (c) linearized potential for the double sine-Gordon model with different values of the parameter $R$.}
   \label{fig_potential}
\end{figure}

Let us consider the double sine-Gordon model described by the following Lagrangian in ($1+1$) dimensions
\begin{equation}
\mathcal{L}=\frac{1}{2}(\partial_\mu\phi)^2-V_R(\phi),
\end{equation}
where $\phi$ is a scalar field and the potential term is given by \cite{campbell1986kink}
\begin{equation}
\label{eq_potential}
V_R(\phi)=\tanh^2R(1-\cos\phi)+\frac{4}{\cosh^2R}\left(1+\cos\frac{\phi}{2}\right).
\end{equation}
The potential is periodic and is shown in Fig.~\ref{fig_potential}(a) for some values of $R$. It is clear from eq.~(\ref{eq_potential}) that for $R=0$ and in the limit $R\to\infty$ the potential approaches the sine-Gordon one with periods $4\pi$ and $2\pi$, respectively. For $\sinh^{-1}(1)\leq R\leq 0$, the potential has only one maximum per period, as in the sine-Gordon model. On the other hand, for  $R>\sinh^{-1}(1)$, a local minimum appears, which approaches zero in the large $R$ limit.

Solving the equation of motion of the system gives the following kink and antikink solutions (see Fig.~\ref{fig_potential}(b))
\begin{equation}
\phi_{k({\bar{k}})}=4\pi n\pm4\tan^{-1}\left(\frac{\sinh x}{\cosh R}\right),
\end{equation}
when the scalar field is static.
If we write the sine-Gordon kink solution as $\phi_{SG}=4\tan^{-1}\exp(x)$, this becomes
\begin{equation}
\phi_{k({\bar{k}})}=4\pi n\pm[\phi_{SG}(x+R)-\phi_{SG}(R-x)].
\end{equation}
Therefore, one may interpret the double sine-Gordon kink as a superposition of two sine-Gordon kinks separated by a distance of $2R$.

We are interested in collisions between double sine-Gordon kinks that are vibrating. This is important to gain a deeper insight into the vibrational or shape modes' role in the resonance windows. Therefore, we look for these modes in the stability equation of the kink solution. Writing $\phi=\phi_k+\psi e^{i\omega t}$, where $\psi$ is a small perturbation, leads to the Schr\"odinger-like equation
\begin{equation}
-\psi^{\prime\prime}+U_R\psi=\omega^2\psi,
\end{equation}
where the linearized stability potential is given by \cite{gani2018scattering}
\begin{equation}
U_R=\frac{8\tanh^2R}{(1+\text{sech}^2R\sinh^2x)^2}+\frac{2(3-4\cosh^2R)}{\cosh^2R(1+\text{sech}^2R\sinh^2x)}+1.
\end{equation}
The potential is shown in Fig.~\ref{fig_potential}(c) for several values of $R$. For large $R$, it clearly splits into two sine-Gordon wells.

\begin{figure}[tbp]
\centering
   \includegraphics[width=0.45\linewidth]{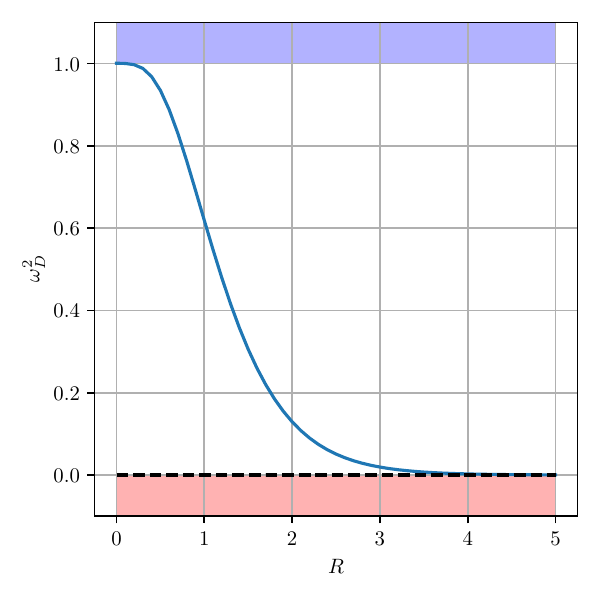}
   \caption{Spectrum of the linearized equation around a kink solution. The solid line is the vibrational mode, and the dashed line is the zero mode. The blue region represents the continuum region.}
   \label{fig_spectrum}
\end{figure}

The spectrum of the linearized equation as a function of $R$ is shown in Fig.~\ref{fig_spectrum}. It contains a zero mode, as required by translational invariance, and a single vibrational mode. The normalized profile of the shape mode is denoted by $\psi_D$ with eigenvalue $\omega_D^2$. It is simple to show that the energy stored in the shape mode with amplitude $A$ is equal to $E_D=\frac{1}{2}\omega_D^2A^2$. 
There is also a continuum of states starting at $\omega^2=1$. For $R=0$ the vibrational mode disappears in the continuum, as the kink solution becomes the sine-Gordon one. For $R\gg1$ the value of $\omega_D^2$ approaches zero as the kink tends to two well-separated sine-Gordon kinks.

\begin{figure}[tbp]
\centering
   \includegraphics[width=\linewidth]{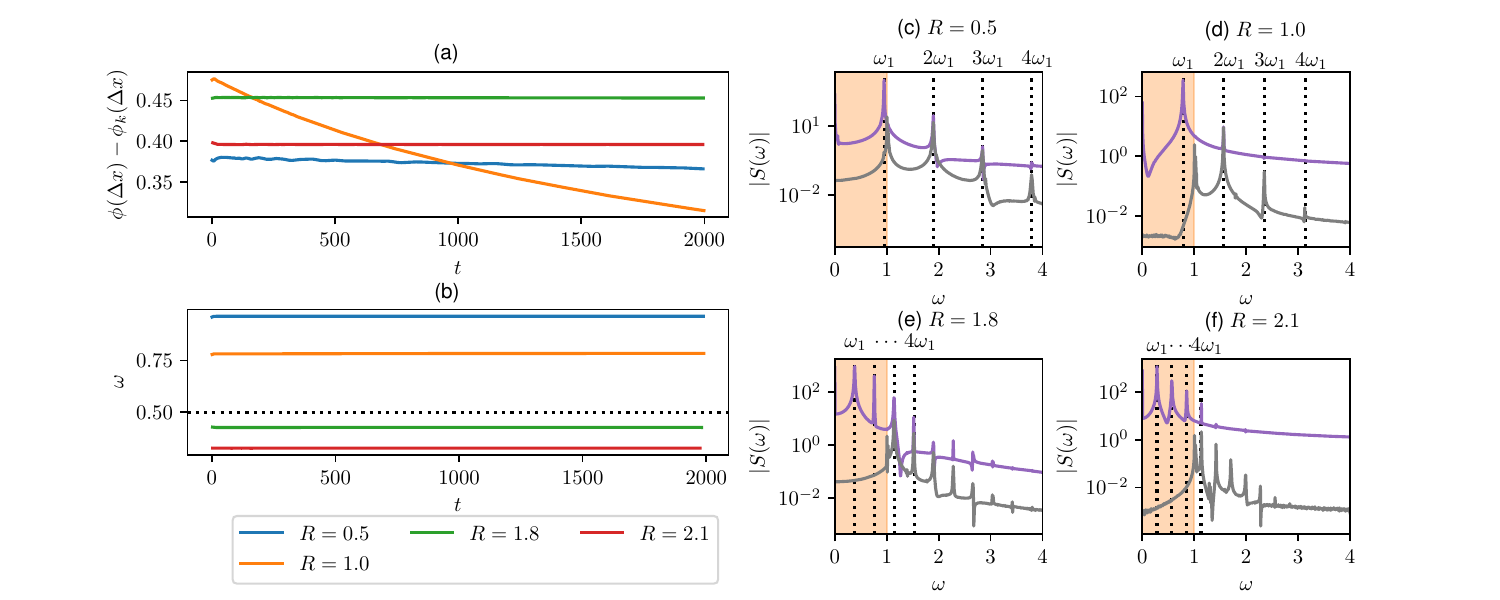}
   \caption{Evolution of the (a) amplitude and (b) frequency of oscillation of the shape mode at $\Delta x=1.465$. (c)-(f) Absolute value of the Fourier transform of the data from (a) and (b), considering $t>1000$. The purple line corresponds to the shape mode vibration at $x=\Delta x$ and the gray line to the far-field vibration at $x=25.0$. The initial amplitude is $A=1.0$.}
   \label{fig_deepmode}
\end{figure}

Before delving into the kink-antikink collision, let us first discuss the behavior of a single wobbling kink. We start with a single excited kink at the origin, and we integrate the equations of motion as described in Appendix \ref{ap1}. It is known that a vibrating kink decays through the coupling to radiation, due to higher-order terms of the wobbling amplitude in the equations of motion, via the harmonics of the wobbling frequency. In the double sine-Gordon model, the decay also follows the Manton-Merabet pattern \cite{manton1997kinks} as shown in Fig.~\ref{fig_deepmode} for $R=0.5$ and $R=1.0$. However, interestingly, the decay for $R=0.5$ is much less pronounced, and this phenomenon becomes more evident approaching the integrable limit. Moving to higher values of $R$, we reach a point where the shape mode frequency becomes lower than the half-threshold value. In this case, the decay occurs through higher harmonics \cite{romanczukiewicz2018oscillons}, as can be seen in the right panels. There, we take the Fourier transform $S(\omega)$ of the field time series near the wobbling kink and far from it and plot its absolute value. We see that the far-field spectrum, which corresponds to the emitted radiation, has peaked in the harmonics of the vibration frequency above the continuum threshold. For $R=1.8$ and $R=2.1$, the decay occurs mainly through the second and third harmonic, respectively, and, therefore, the perturbation is much more stable \cite{romanczukiewicz2018oscillons}, as can be seen in the amplitude evolution. 

\section{Collision simulations}
\label{simul}

To initialize the collision we take the following configuration
\begin{equation}
\label{eq_ansatz}
\phi=\phi_k\left(\xi_+\right)-\phi_k\left(\xi_-\right)-2\pi+A\sin(\omega_D \tau_+)\psi_D\left(\xi_+\right)-A\sin(\omega_D \tau_-)\psi_D\left(\xi_-\right),
\end{equation}
evaluated at $t=0$. For the time evolution we defined the boosted coordinates $\xi_\pm=\gamma(x\pm x_0\mp v_it)$ and $\tau_\pm=\gamma(t\mp v_ix)$\footnote{It is a common mistake to only impose the boost on $x$ coordinate and neglect it in the time coordinate.}, where $\gamma=1/\sqrt{1-v_i^2}$. It consists of a kink and an antikink separated by a distance $2x_0$, fixed at the value $2x_0=30$. They approach each other with velocity $v_i$ and $-v_i$, respectively, and at the same time, they wobble with the bound frequency $\omega_D^2$ and amplitude $A$. For $A=0$, this leads to the usual kink-antikink collision \cite{campbell1986kink}. The technical details of the simulations are described in Appendix \ref{ap1}.

From eq.~(\ref{eq_ansatz}), it is clear that the wobbling oscillation frequency is Lorentz contracted in the center of mass frame. If we measure the time evolution of the field at a point moving with the kink in the form $x=v_it+\alpha$, the wobbling term becomes
\begin{equation}
\label{eq_contrac}
A\sin\left(\frac{\omega_D}{\gamma}t-\gamma\omega_Dv_i\alpha\right)\psi_D(\gamma(x_0+\alpha)),
\end{equation}
which oscillates with a Lorentz contracted frequency $\omega_D/\gamma$. This transverse Doppler effect was also shown to appear in a more detailed description of wobbling kinks investigated in \cite{barashenkov2009wobbling}. We will observe the same effect after the collision, where the frequency is contracted according to the system's final velocity.

\begin{figure}[tbp]
\centering
   \includegraphics[width=\linewidth]{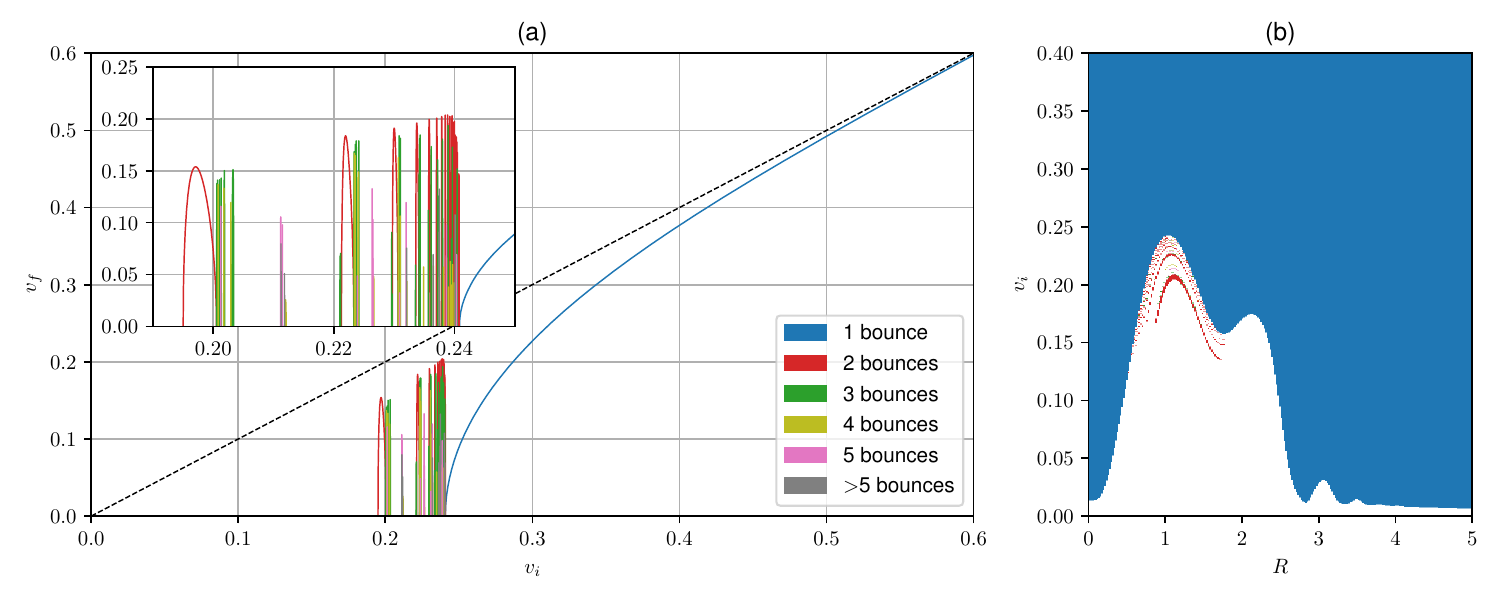}
   \caption{(a) Final velocity as a function of the initial velocity for kink-antikink collisions, considering $R=1.0$. (b) Number of bounces before escape for kink-antikink collisions as a function of the initial velocity $v_i$ and $R$. For both diagrams, we set $A=0.0$.}
   \label{fig_resonance2}
\end{figure}

The double sine-Gordon model with $A=0$ exhibits resonance windows in a finite range of the parameter $R$. This occurs presumably because, for both very small and very large $R$, the model becomes the sine-Gordon one, which is integrable and does not have resonance windows. We show the final velocity as a function of the initial velocity for a specific value of $R=1.0$ in Fig.~\ref{fig_resonance2}(a). This value is in the region where the system exhibits resonance windows. The fractal structure is similar to the $\phi^4$ model. However, there is a significant difference. Due to the periodicity of the potential, the kink and the antikink may either cross or reflect. An even number of bounces means that the kinks reflected, while an odd number means that the kinks crossed.

In Fig.~\ref{fig_resonance2}(b), we show the number of bounces as a function of $R$ and $v_i$, the initial velocity of the incoming kink-antikink. The colors represent the number of bounces as indicated in Fig.~\ref{fig_resonance2}(a). The white region is where the kink and the antikink annihilate. In the blue region, separation occurs after a single bounce. The interface between white and blue determines the critical velocity value, which agrees with the non-monotonic behavior reported in \cite{gani2018scattering}. Moreover, the resonance domain with a fractal structure, the region with dots in red and other colors, is located approximately in the interval $0.52\leq R\leq 1.78$. This interval translates to a frequency interval $0.437<\omega_D<0.963$. Of course, the actual resonance domain would be slightly larger because the resonance windows become increasingly narrower and consequently more difficult to locate. 

\begin{figure}[tbp]
\centering
   \includegraphics[width=\linewidth]{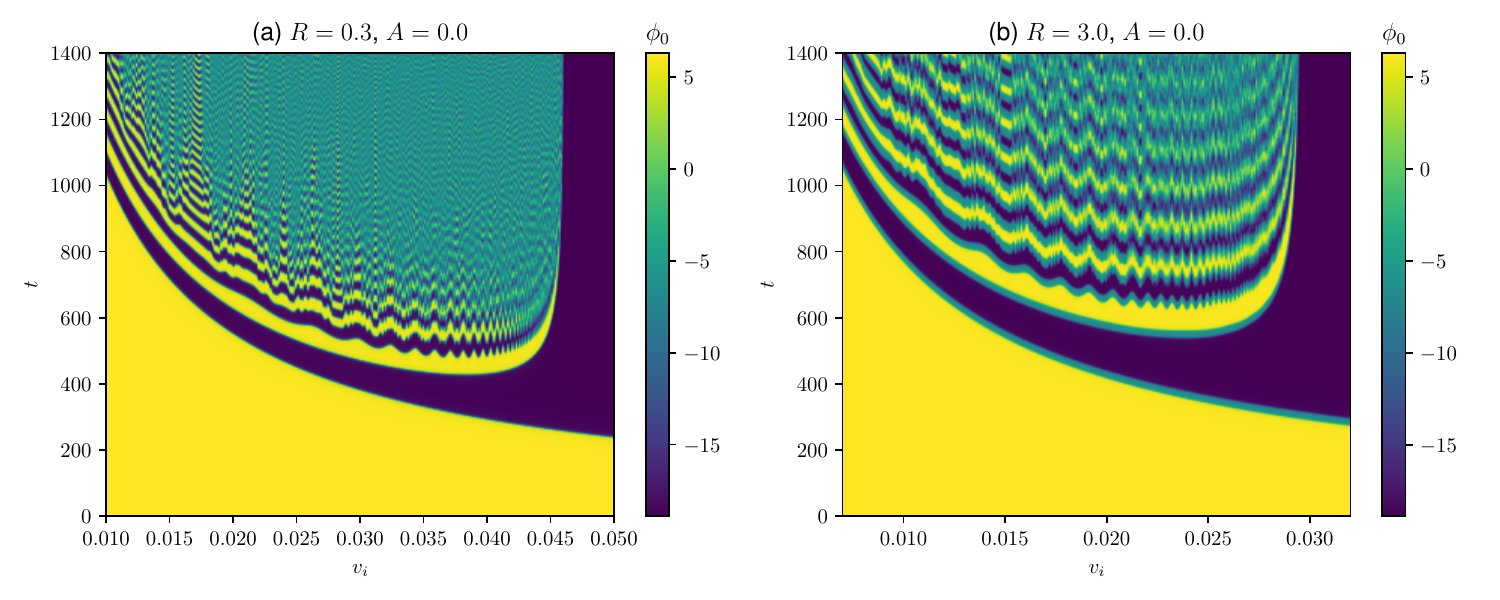}
   \caption{The value of the field at the center of collision as a function of time and the initial velocity.}
   \label{fig_fieldcenter}
\end{figure}

To confirm that the model does not have resonance windows for large and small $R$, we plot the time evolution of the field at the collision center as a function of the initial velocity. This is shown in Fig.~\ref{fig_fieldcenter} for $R=0.3$ and $R=3.0$. In both cases, we can only see false resonance windows, despite the presence of a vibrational mode. It means that the existence of the vibrational mode does not guarantee the appearance of a fractal structure. Therefore, the vibrational mode is not sufficient for the resonant energy exchange mechanism. In \cite{simas2020solitary}, the authors obtained a similar result near the integrable limit of a case of the double sine-Gordon different from what is investigated here and, in \cite{dorey2021resonance}, the authors also found false resonance windows near the integrable regime of a deformed sine-Gordon model. We searched carefully for resonance windows with precision $\Delta v_i=10^{-5}$ and no resonance windows were found.

\begin{figure}[tbp]
\centering
   \includegraphics[width=\linewidth]{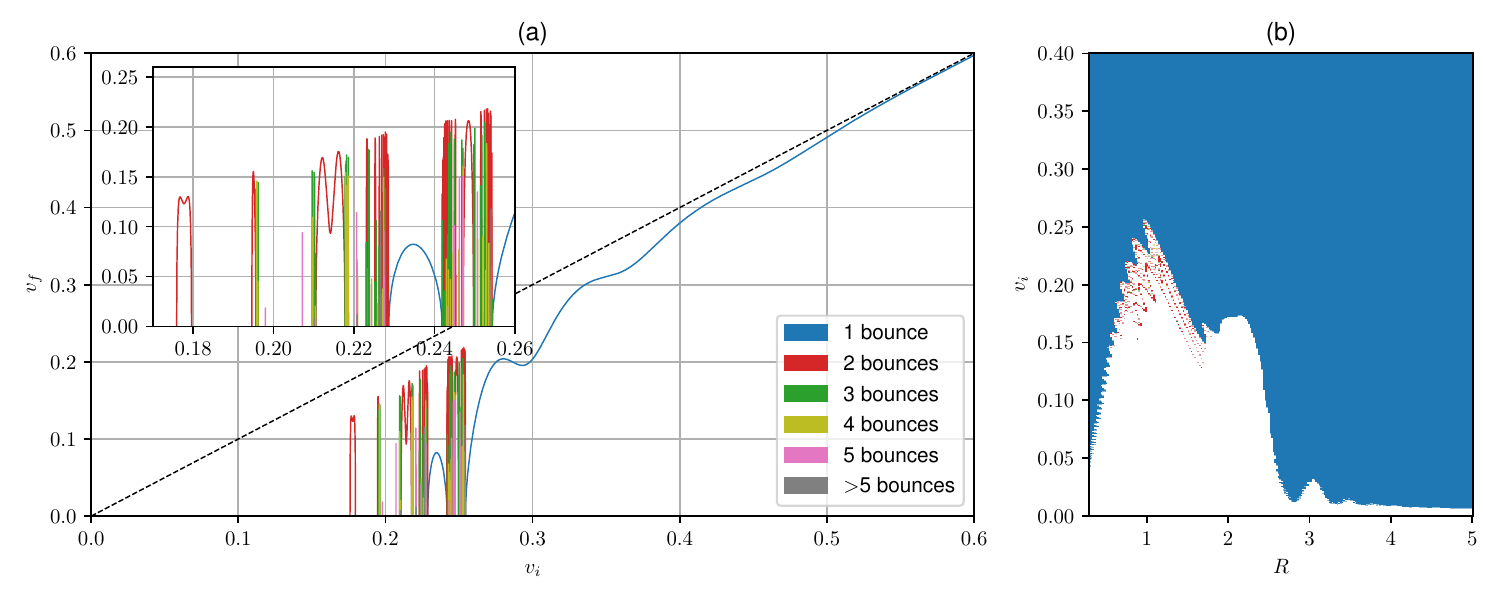}
   \caption{(a) Final velocity as a function of the initial velocity for kink-antikink collisions, considering $R=1.0$. (b) Number of bounces before escape for kink-antikink collisions as a function of the initial velocity $v_i$ and $R$. For both diagrams, we set $A=0.1$.}
   \label{fig_resonance3}
\end{figure}

Now we would like to see how the result with $A\neq0$ compares with the previous one. We start with a small value of $A=0.1$. The final velocity dependence on the initial one is shown in Fig.~\ref{fig_resonance3}(a) for $R=1.0$. We observe that the blue crossing curve, which indicates separation after one-bounce, now oscillates as we vary $v_i$. This occurs because, for different velocities, the phase of the wobbling at the collision varies. Before the collision, the wobbling amplitude evolves approximately as
\begin{equation}
\label{eq_st}
S(t)=A\exp\left[i\left(\frac{\omega_D}{\gamma}t+\theta_0\right)\right],
\end{equation}
where $\theta_0$ is an initial constant. Moreover, the collision occurs approximately at $t=x_0/v_i$. Therefore, the dependence of the phase $\theta$ at the collision on $v_i$ can be estimated as
\begin{equation}
\label{eq_phase}
\theta=\sqrt{1-v_i^2}\, \frac{\omega_Dx_0}{v_i}+\theta_0.
\end{equation}
Crossing after one bounce always occurs after a critical velocity. However, the oscillation in the amplitude, shown in eq.~(\ref{eq_st}), causes the one-bounce crossing curve to split, creating one or more isolated one-bounce windows. Near these windows, there appears a nested structure of higher-bounce windows. The gap between the two one-bounce regions occurs because, in this region, the wobbling is out of phase at the collision time. Similarly, it is also possible to see oscillating behavior in some two-bounce windows. This novel phenomenon was also observed in the collision between wobbling kinks of the $\phi^4$ model \cite{izquierdo2021scattering}.

In Fig.~\ref{fig_resonance3}(b), we summarize the behavior of the system for both dependences on $R$ and $v_i$. One important detail is that in contrast with the case of $A=0$, for small $R$ it becomes difficult to find the shape mode profile numerically with enough precision as it approaches the continuum. That is the reason we have shown the figure starting with a small nonzero $R$.
It is possible to see that the one-bounce crossing region oscillates with many spines in double sine-Gordon model, similar to what was described in \cite{dorey2021resonance}\footnote{The term spine was coined in \cite{dorey2021resonance}.}. These spines originate from the formation of one-bounce resonance windows as explained in detail in \cite{dorey2021resonance}. In short, the formation of spines can be pictured by tracking how the one-bounce windows appear, move and connect as we change $R$. Another interesting point is that the wobbling energy is enough to create one-bounce windows even for the values of $R$ where there was no fractal structure in the absence of wobbling, $A=0$.

\begin{figure}[tbp]
\centering
   \includegraphics[width=\linewidth]{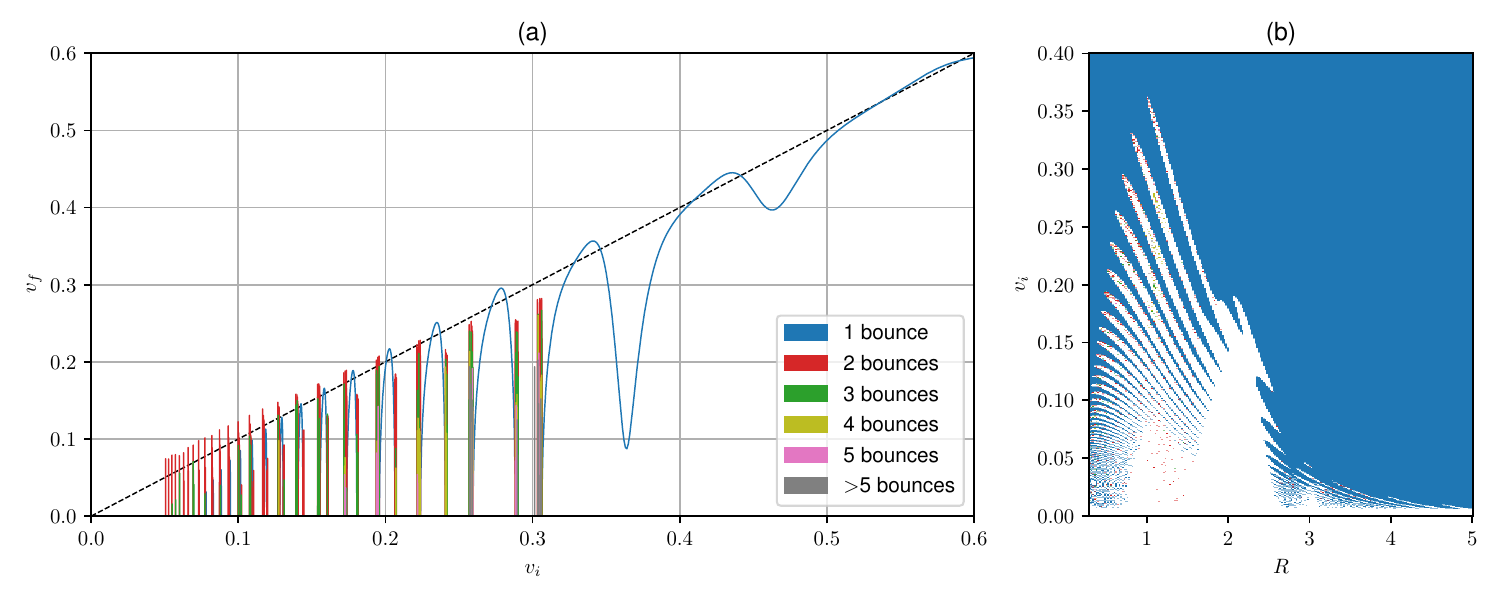}
   \caption{(a) Final velocity as a function of the initial velocity for kink-antikink collisions, considering $R=1.0$. (b) Number of bounces before escape for kink-antikink collisions as a function of the initial velocity $v_i$ and $R$. For both diagrams, we set $A=0.8$.}
   \label{fig_resonance4}
\end{figure}

For $A=0.1$, the wobbling effect is small, while if we increase $A$, it becomes more pronounced. Figure~\ref{fig_resonance4}(a) shows the system's behavior for a large wobbling amplitude $A=0.8$. As one can see, there is an oscillating pattern in the curve of $v_f$ as a function of $v_i$, as well as many isolated one-bounce resonance windows. This occurs in two ways. The first one is the process we described before. The one-bounce crossing curve is split in a region where the wobbling phase is such that the translational energy is lost at the collision. The second way is that a one-bounce window appears in a place where previously there was not any window. These windows show up because the energy transferred from the wobbling at the collision is now high enough, letting the kinks separate. This phenomenon was also reported in \cite{izquierdo2021scattering} for the $\phi^4$ model.
Interestingly, near the boundary of a one-bounce window, there is a nested structure of higher-bounce windows. Furthermore, two essential features should be noticed. First, as we decrease $v_i$, the one-bounce resonance windows disappear, we see a pattern of two-bounce resonance windows that continues all the way down to zero initial velocity. Second, the final velocity may become higher than the initial one in some regions. Both features occur because the vibrational energy is considerable and, therefore, there is a large amount of energy available that may turn into translational energy at the collision. 

The system's behavior for both dependences on $R$ and $v_i$, although now with a larger wobbling amplitude $A=0.8$, is summarized in Fig.~\ref{fig_resonance4}(b). The figure shows an intricate structure of one-bounce resonance windows with many spines in the blue region. This pattern shows that many one-bounce resonance windows appear for large values of $A$ in the two ways previously described. In particular, we observe that the critical velocity after which one-bounce crossing always occurs is larger than the case with a smaller value of $A$. Due to the large amplitude of wobbling, when it is out of phase, there appear annihilation windows even for initial velocities as large as $v_i=0.35$. These annihilation windows with large initial velocities are precisely where the final velocity curve for one-bounce crossing splits. Interestingly, we  observe that one-bounce resonance windows spines appear in the whole region of $R$ that we considered.

\begin{figure}[tbp]
\centering
   \begin{subfigure}[b]{\textwidth}
         \centering
         \includegraphics[width=\textwidth]{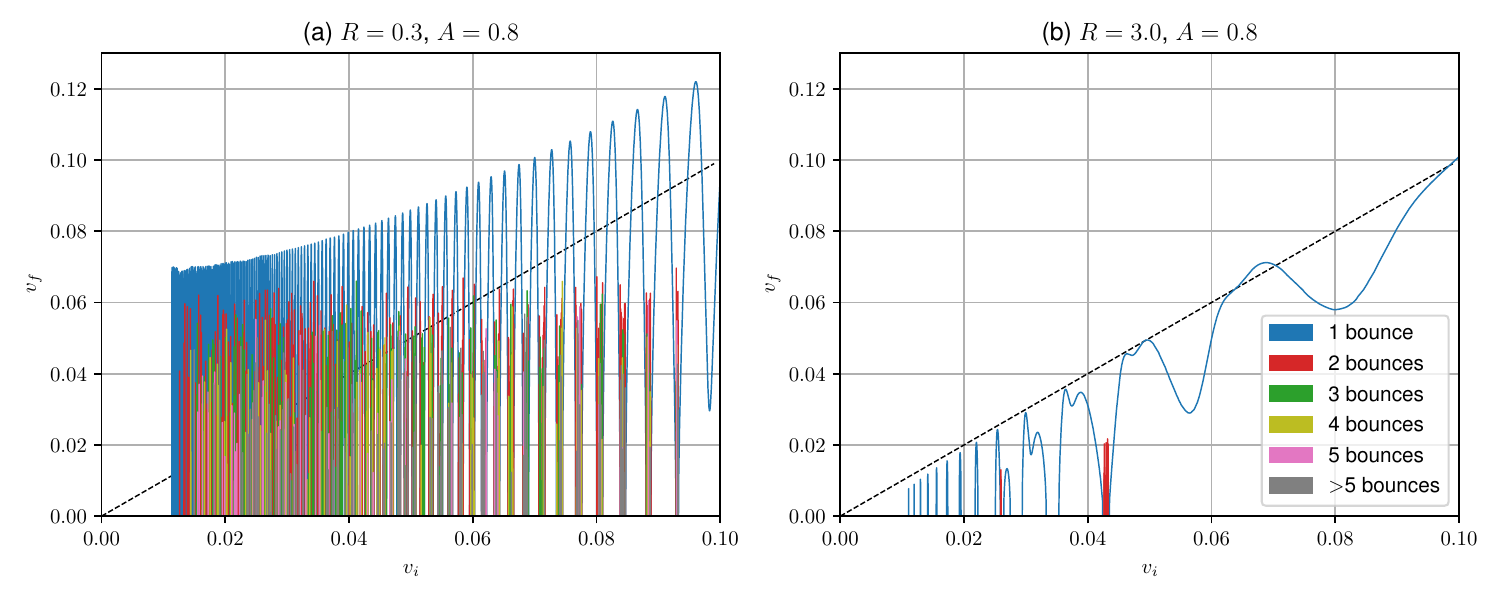}
     \end{subfigure}
     \begin{subfigure}[b]{\textwidth}
         \centering
         \includegraphics[width=\textwidth]{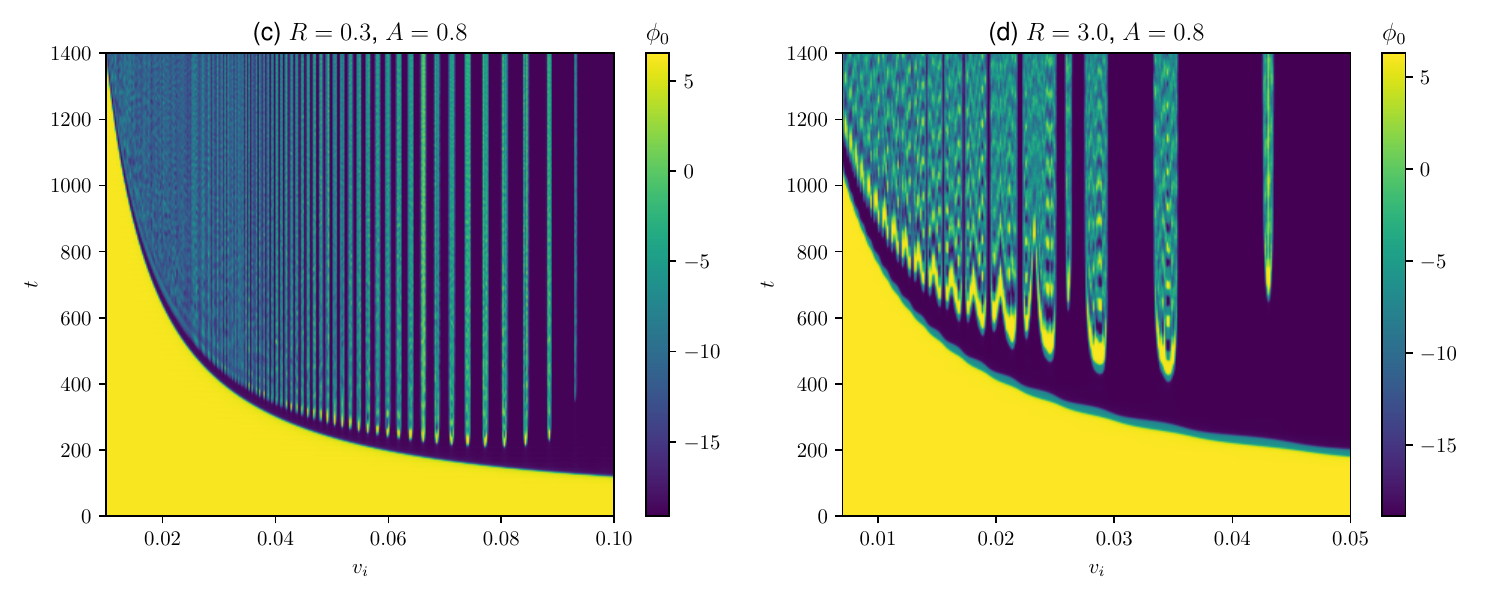}
     \end{subfigure}
   \caption{(a) and (b) Final velocity as a function of the initial velocity for kink-antikink collisions. (c) and (d) Value of the field at the center of collision as a function of time and the initial velocity .}
   \label{fig_small_and_large}
\end{figure}

Now, let us investigate the existence of higher-bounce windows for small and large $R$ when the wobbling is turned on. The result is shown in Fig.~\ref{fig_small_and_large}, where we fix again $R=0.3$ and $R=3.0$. In the upper panels, we plot the final velocity as a function of the initial one. For $R=0.3$, the wobbling frequency is large and the critical velocity is small. In this case, we can see from eq.~(\ref{eq_phase}) that the wobbling phase varies much faster with $v_i$ and many one-bounce windows appear because the phase quickly alternates between constructive and destructive energy exchange. Moreover, there are many higher-bounce resonance windows at the border of these windows, contrary to the $A=0$ case for the same value of $R$. For $R=3.0$, there are much fewer one-bounce windows because the wobbling frequency is much smaller than the $R=0.3$ case. We still find two-bounce windows in the neighborhood of the one-bounce ones, albeit not as many as the $R=0.3$ case. It can be justified by the fact that the wobbling energy is proportional to $\omega_D^2$, making it much smaller for $R=3.0$. As for both small and large $R$ the resonance is recovered for sufficiently large amplitude $A$, one could argue that the system possesses a hidden resonance structure before the wobbling is turned on.  

We also plot the field at the center of the collision as a function of the initial velocity in the lower panels of Fig.~\ref{fig_small_and_large}, for comparison. One can clearly see the structure of one-bounce resonance windows, but it is not possible to see higher-bounce ones in the resolution of the figure because they are extremely narrow.
Due to the near threshold wobbling frequency of the $R=0.3$ case, it could be a good candidate for exhibiting the spectral phenomenon reported in \cite{adam2019spectral, adam2020kink}. However, we did not find this phenomenon for double sine-Gordon model. We think this is due to the fact that the interkink attractive force is larger than the potential spectral wall making it impossible to isolate the effect.

\begin{figure}[tbp]
\centering
   \includegraphics[width=\linewidth]{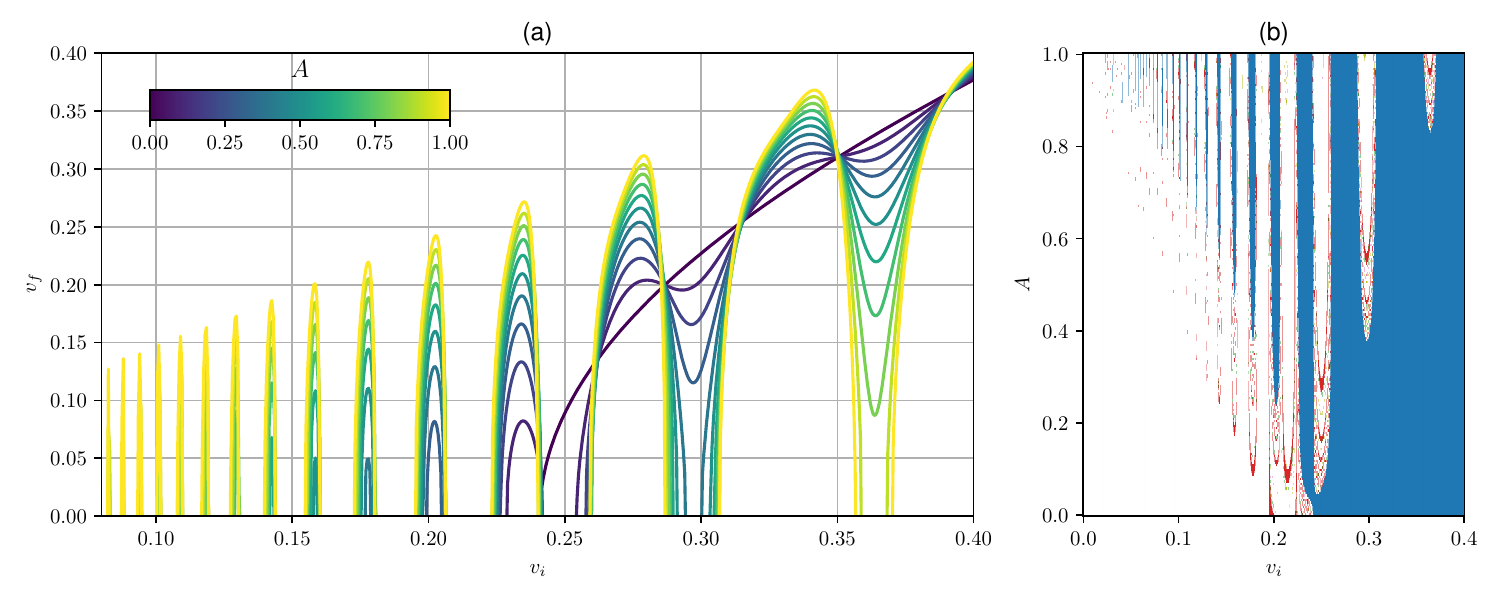}
   \caption{(a) Final velocity as a function of the initial velocity for kink-antikink collisions for different values of $A$. We fix $R=1.0$. (b) Number of bounces before escape for kink-antikink collisions as a function of the initial velocity $v_i$ and $A$.}
   \label{fig_resonance5}
\end{figure}

The one-bounce windows' behavior for the range $0\le A \le 1$ and fixed value of $R=1.0$ is summarized in Fig.~\ref{fig_resonance5}(a). This is the best way to visualize the two processes of the creation of one-bounce windows. It is possible to see that gradually, as $A$ is increased, the initial one-bounce curve splits, and new ones start to emerge. If $A$ is large enough, whenever the vibration is in phase at collision, there are one-bounce resonance windows, while there are annihilation windows otherwise.
A more detailed structure of n-bounce windows as a function of the system's relevant parameters is shown in Fig.~\ref{fig_resonance5}(b). As one can see, the critical velocity and the number of isolated one-bounce windows increase as the amplitude of oscillation increases, consistent with Fig.~\ref{fig_resonance5}(a) along with our arguments before. 
Moreover, the figure shows that the system also exhibits higher-bounce windows. However, the higher-bounce windows pattern is rather scarce because the higher-bounce resonance windows are narrow for the double sine-Gordon model. Nevertheless, it is also possible to see some higher-bounce resonance windows near the boundary of one-bounce resonance windows and in the region where $A$ is large and $v_i$ is small. In fact, it is easier to visualize this phenomenon in Fig.~\ref{fig_resonance4} for $A=0.8$ where higher-order bounce windows accumulate in these two regions. Furthermore, the alternation between one-bounce and annihilation windows is clearly shown in Fig.~\ref{fig_resonance5}(b) when $A$ is large. This alternation depends on the wobbling phase right before the first collision, given by eq.~(\ref{eq_phase}).

\begin{figure}[tbp]
\centering
   \includegraphics[width=0.5\linewidth]{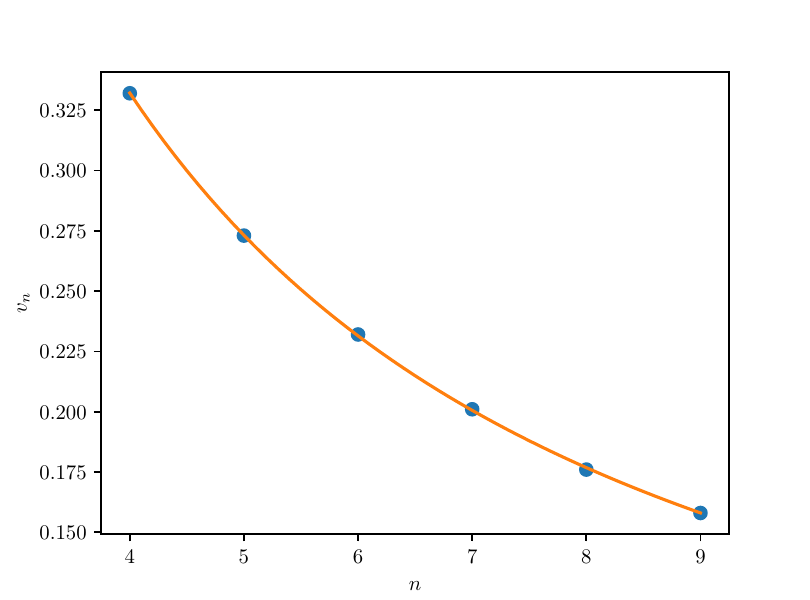}
   \caption{One-bounce windows center $v_n$ as a function of the integer $n$ for $R=1.0$. The solid curve is a fit of the type given in eq.~(\ref{fig_fit}). }
   \label{fig_vn}
\end{figure}

We can find an approximate expression for the location of the one-bounce windows using the method proposed by Campbell et al. \cite{campbell1983resonance,campbell1986kink}. Assuming that the amplitude is small and the system is invariant under time reversal, the authors found that the wobbling amplitude after the collision $S^\prime$ is approximately given by
\begin{equation}
\label{eq_sprime}
S^\prime=-\frac{\rho}{\rho^*}S+\rho,
\end{equation}
where $\rho$ is a complex constant that depends on the critical velocity. The center of the one-bounce windows should be located approximately at the point where the amplitude of wobbling is minimum. Plugging eq.~(\ref{eq_phase}) in eq.~(\ref{eq_sprime}) we find 
\begin{equation}
S^\prime=-A\exp\left[i\left(\sqrt{1-v_i^2}\frac{\omega_Dx_0}{v_i}+\theta_0+2\theta_\rho\right)\right]+|\rho|\exp(i\theta_\rho),
\end{equation}
where $\theta_\rho$ is the argument of $\rho$. The amplitude $S^\prime$ has a minimum absolute value for initial velocities $v_n$ such that
\begin{equation}
\sqrt{1-v_n^2}\frac{\omega_Dx_0}{v_n}+\theta_0+2\theta_\rho=2\pi n+\theta_\rho
\end{equation}
for some integer $n$. More explicitly we have
\begin{equation}
v_n=\frac{\omega_Dx_0}{\sqrt{(2\pi n-\delta)^2+\omega_D^2x_0^2}},
\end{equation}
where $\delta\equiv \theta_\rho+\theta_0$. Notice that the location of the centers is almost independent of the amplitude $A$, as can be observed in Fig.~\ref{fig_resonance5}. Therefore, we choose $A=1.0$ to measure the center of the one-bounce windows because there are more windows for this value.  Figure~\ref{fig_vn} shows the centers of the one-bounce windows as a function of an integer $n$ for $R=1.0$. Fitting a curve of the type
\begin{equation}
\label{fig_fit}
v_n=\frac{a}{\sqrt{(2\pi n-b)^2+a^2}},
\end{equation}
where $a$ and $b$ are fitting parameters, we find $a\simeq 9.22$, which should be compared to the theoretical value $\omega_Dx_0\simeq11.80$. The agreement is not remarkable, which is expected given all the approximations in the equations. However, the above analysis gives a correct qualitative picture of the exact, within the numerical precision, numerical simulations.

\begin{figure}[tbp]
\centering
   \includegraphics[width=\linewidth]{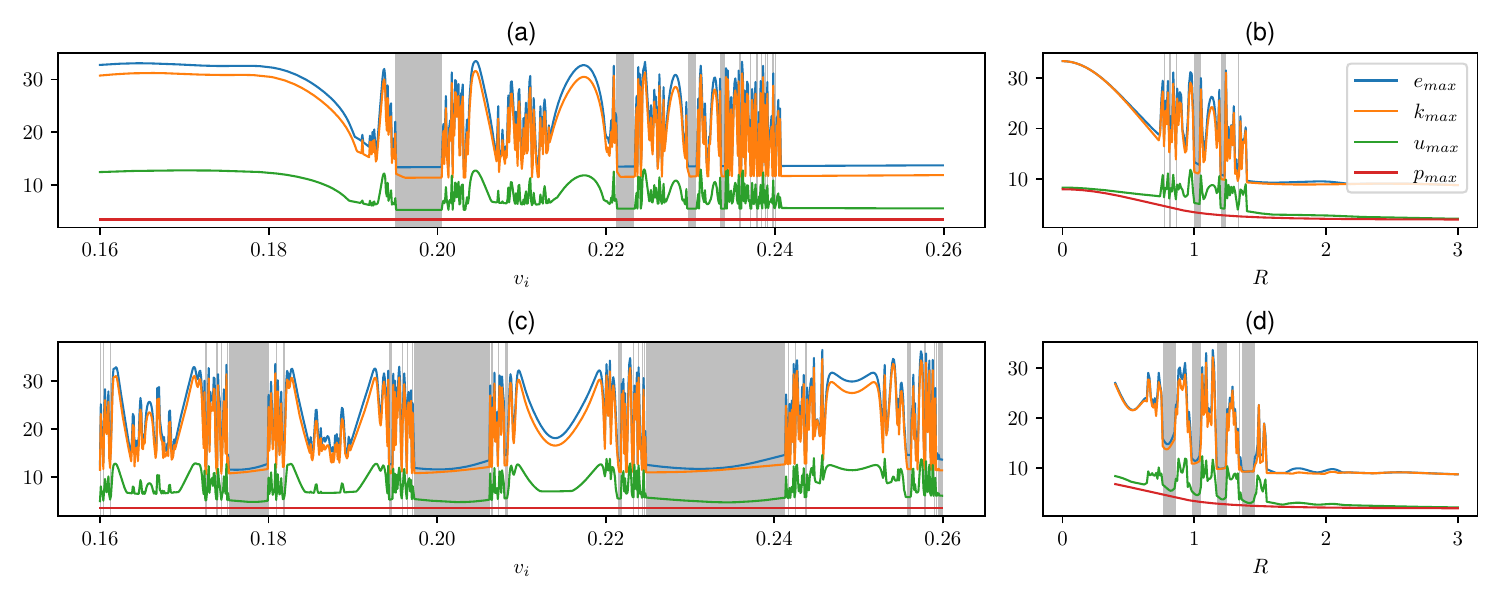}
   \caption{ Maximum energy densities as a function of (a) $v_i$ taking  $R=1.0$ (b) $R$ taking $v_i=0.2$, both for $A=0.0$. (c) and (d) same quantities now with $A=0.5$. The marked regions consist of one-bounce and two-bounce resonance windows.}
   \label{fig_maxdens}
\end{figure}

Let us also study the maximum energy densities as a function of the system's parameters as reported in \cite{gani2019multi,zhong2020collision}. The energy density is given by $e=k+u+p$ where the three terms in this expression are the kinetic $k=\frac{1}{2} (\partial_t \phi)^2$, gradient $u=\frac{1}{2} (\partial_x \phi)^2$ and potential $p=V(\phi)$ energy densities, respectively. During the collision, each of them will reach a maximum value, which we denote by a subscript $max$, at some position in spacetime. Figure~\ref{fig_maxdens} shows the maximum energy densities as a function of $v_i$ and $R$ in both scenarios with and without wobbling. In all the figures, we observe that, interestingly, the maximum energy densities vary smoothly inside the resonance windows, as shown in the highlighted one-bounce and two-bounce regions. This behavior is in clear contrast with erratic behavior in most points outside the resonance windows related to the bion formation, which is known to evolve chaotically \cite{anninos1991fractal}. The result in Fig.~\ref{fig_maxdens}(b) matches the analogous figure in reference \cite{gani2019multi}. The double sine-Gordon is well known to form long-lived bound states, called oscillons, between the sine-Gordon subkinks. As far as we checked, the dependence of the maximum energy density on the initial velocity is also smooth in the regions where there is oscillon formation.

\begin{figure}[tbp]
\centering
   \includegraphics[width=0.7\linewidth]{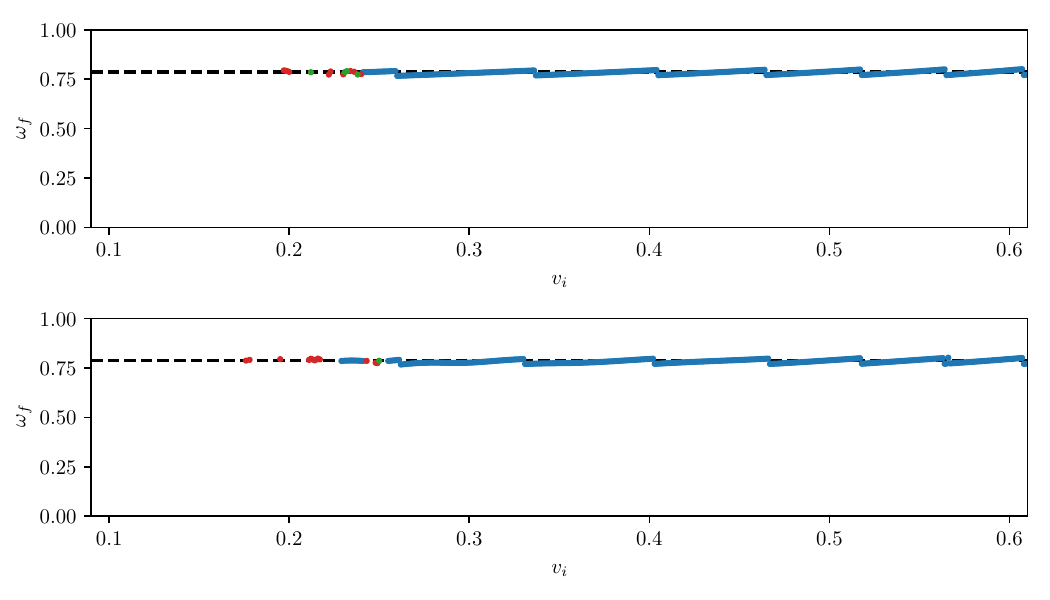}
   \caption{  Final wobbling frequency as a function of the initial velocity for $A=0.0$ (top) and $A=0.1$ (bottom) with $R=1.0$. The frequency was measured in the reference frame with the kink at rest, where the Lorentz contraction factor is compensated. The dashed line is the theoretical frequency of the shape mode $\omega_D\simeq 0.7867$. The color scheme represents the number of bounces as in previous figures.}
   \label{fig_freq}
\end{figure}

Finally, we would like to conclude with the measurement of the final wobbling frequency $\omega_f$, the wobbling frequency of the system after the collision. We measure the final frequency by taking the field's value at a distance from the center, where the amplitude of wobbling has a maximum theoretical value. Then we subtract the kink contribution from the field and take the Fast Fourier Transform of the evolution of this value. The frequency with the largest amplitude in the power spectrum is plotted in Fig.~\ref{fig_freq}. In the figure, we divide the obtained frequency by $\sqrt{1-v_f^2}$ to compensate the Lorentz contraction factor and change to the frame where the kink is at rest, as explained before. It is clear from the figure that, in this frame, the kink indeed wobbles at the shape mode's theoretical frequency $\omega_D$. This result serves as a nice consistency check and is similar to what was obtained in \cite{izquierdo2021scattering} for the $\phi^4$ model. Moreover, higher-order effects in the amplitude can change the wobbling frequency from the lowest order value $\omega_D$ \cite{barashenkov2009wobbling}. However, this effect is negligible compared with the Lorentz contraction one, consistent with the results in Fig.~\ref{fig_freq}.

\section{Conclusion}

In this paper, we started with a subset of the double sine-Gordon model that depends on the parameter $R$. This model is well-known and coincides with the sine-Gordon model for $R=0$ and in the limit $R\to\infty$. The model has at maximum one shape mode in the whole range of parameters, which is important for the appearance of resonance windows. The shape mode decays into radiation through the first or higher harmonics depending on the value of the oscillating frequency and is more stable in the latter case.
Then, we investigated collisions between the kink and antikink in this model. The collisions exhibit the familiar fractal structure of resonance windows. However, it only exists when the system is far from the integrable limits, despite having a shape mode. It implies that the existence of a shape mode is not a sufficient condition for the resonance energy exchange mechanism. Furthermore, we observed the previously reported non-monotonic critical velocity dependence on $R$. 

Next, we added wobbling to the kink and antikink and studied collisions again. We found that the final velocity depends on the wobbling phase in an oscillatory way. If the wobbling is in phase in the first collision, we may observe a one-bounce crossing. This gives rise to separation after a critical velocity and also to isolated one-bounce windows, which do not occur in the absence of wobbling. On the other hand, if the wobbling is out of phase, it causes annihilation or the formation of higher-bounce resonance windows. This phenomenon is even more pronounced for larger wobbling amplitudes, where we clearly see an alternating structure of these two behaviors. This result is similar to what was reported in \cite{izquierdo2021scattering} for the $\phi^4$ model. We found one-bounce windows in the whole range of $R$, which appears as an intricate structure of spines in the reflection region. Interestingly, we found that when the wobbling is turned on, the fractal structure is gradually recovered in the region with small and large $R$.

In the double sine-Gordon model, every bounce corresponds to the kink-antikink crossing instead of reflecting, in contrast with the $\phi^4$ model. We could show that the one-bounce windows' peak has a simple dependence on the wobbling phase. It can be found approximately considering a linear relation between the wobbling amplitude before and after the collision.
We also measured the wobbling frequency of the kinks after the collision and found that indeed the kink wobbles at the theoretical value of the frequency $\omega_D$ once you change to the frame where the kink is at rest. This Lorentz contraction is manifest in our equation for a boosted wobbling kink.

The maximum energy density in the kinks collisions is an interesting way of studying the fractal structure of resonance windows \cite{gani2019multi,zhong2020collision}. Curiously, we were able to show that it also works for our system. The energy density shows smooth behavior in the windows, independent of the system's parameters, despite the chaotic structure of the collision. As far as we understand, the erratic behavior outside the windows is related to the bion formation. Moreover, in some of the smooth regions outside the resonance windows, we observed oscillon formation. However, to conclude that this is indeed the case whenever there is an oscillon formation needs a more thorough investigation.

This study may shed light on the mechanism of resonance windows formation, which is approximately described by the energy exchange mechanism of Campbel et al. This mechanism has finally received a compelling quantitative confirmation in a recent work \cite{manton2021collective}, where the authors found appropriate moduli space coordinates \cite{manton2021kink}. 
It would be interesting to see if the reduced model with collective coordinates can reproduce the wobbling kink novel results in a future work. Moreover, our work clearly shows that the regions close to integrable limits in the double sine-Gordon model are not well understood and should be more carefully investigated, possibly with some perturbative method.

\appendix

\section{Numerical technique}
\label{ap1}

To solve the equations of motion numerically, we discretize the space in the interval $-100.0<x<100.0$ using $N=2048$ gridpoints and periodic boundary conditions. The space derivative is done using a Fourier spectral method \cite{trefethen2000spectral} and the time evolution integrated using a 5th order Runge-Kutta method with error control \cite{dormand1980family} implemented with the odeint package in C$++$ \cite{ahnert2011odeint}. We evolve the system until the time $t=1400.0$. Hence, we cannot find resonance windows for very small velocities since there is not enough time for the kinks to separate. The radiation at the boundaries was absorbed by including a damping term in the region $x<-80$ and $x>80.0$. The damping is proportional to a bump function with a maximum value of $5$ and is exactly zero outside this region \cite{christov2020kink}.
A larger box was used to compute the final wobbling frequency to allow for a longer time series and obtain a more precise frequency measurement. The profile of the shape mode $\psi_D(x)$ was computed using the NDEigensystem method in Mathematica. 

\section*{Acknowledgments}

We acknowledge financial support from the Brazilian agencies CAPES and CNPq. AM also thanks financial support from Universidade Federal de Pernambuco Edital Qualis A. We thank Mauro Copelli for the access to his lab's computer cluster, which was essential for conducting the research reported in this paper. We would also like to thank Andrzej Wereszczyński for the fruitful discussions which led to important results.

\end{document}